# Chatbots and ChatGPT: A Bibliometric Analysis and Systematic Review of Publications in Web of Science and Scopus Databases

Hamed Khosravi[1, *], Mohammad Reza Shafie[2], Morteza Hajiabadi[3], Ahmed Shoyeb Raihan[4], Imtiaz Ahmed[5]

[1]*Department of Industrial & Management Systems Engineering, West Virginia University, Morgantown, WV 26505, hk00024@mix.wvu.edu*

[2]*Department of Electrical Engineering, Iran University of Science and Technology, Tehran, Iran, Mr.shafie7731@gmail.com*

[3]*School of Computer Engineering, Iran University of Science and Technology, Tehran, Iran, hajiabadi1377@gmail.com*

[4]*Department of Industrial & Management Systems Engineering, West Virginia University, Morgantown, WV 26505, ar00065@mix.wvu.edu*

[5]*Department of Industrial & Management Systems Engineering, West Virginia University, Morgantown, WV 26505, imtiaz.ahmed@mail.wvu.edu*

[*]*Corresponding Author: hk00024@mix.wvu.edu*

**Abstract**

This paper presents a bibliometric analysis of the scientific literature related to chatbots, focusing specifically on ChatGPT. Chatbots have gained increasing attention recently, with an annual growth rate of 19.16% and 27.19% on the Web of Sciences (WoS) and Scopus, respectively. In this study, we have explored the structure, conceptual evolution, and trends in this field by analyzing data from both Scopus and WoS databases. The research consists of two study phases: (i) an analysis of chatbot literature and (ii) a comprehensive review of scientific documents on ChatGPT. In the first phase, a bibliometric analysis is conducted on all published literature, including articles, book chapters, conference papers, and reviews on chatbots from both Scopus (5839) and WoS (2531) databases covering the period from 1998 to 2023. An in-depth analysis focusing on sources, countries, authors' impact, and keywords has revealed that ChatGPT is the latest trend in the chatbot field.

Consequently, in the second phase, bibliometric analysis has been carried out on ChatGPT publications, and 45 published studies have been analyzed thoroughly based on their methods, novelty, and conclusions. The key areas of interest identified from the study can be classified into three groups: artificial intelligence and related technologies, design and evaluation of conversational agents, and digital technologies and mental health. Overall, the study aims to provide guidelines for researchers to conduct their research more effectively in the field of chatbots and specifically highlight significant areas for future investigation into ChatGPT.

## 1. Introduction

Artificial intelligence (AI) has emerged as a revolutionary technology capable of learning from data and performing tasks that normally require human intelligence (McCarthy, 2007). Due to its ability to develop intelligent systems, AI has been applied in a number of fields (Amara, Hadj Taieb, and Ben Aouicha, 2021). Generate artificial intelligence (GAI) is one type of AI that creates new content instead of identifying or classifying existing content (Jovanovic & Campbell, 2022). In recent years, deep learning, another emerging branch of AI, has been expanded to include a wide range of network structures. Several applications of deep learning can be found today, including computer vision, speech recognition, and natural language processing (NLP) (Chai et al., 2021). NLP is an important component of GAI, as it enables machines to comprehend and generate human language (Aydin & Erdem, 2022). In this technique, words are broken down into smaller components, their relationships are analyzed, and their combinations are explored to form meaningful patterns from their combinations (IBM, 2017; Khodadadi, Ghandiparsi, and Chuah, 2022).

Chatbots are AI applications that simulate human conversation and provide automated responses to user inquiries utilizing NLP (Dwivedi et al., 2021). Chatbots have gained popularity in recent years because they can improve customer service, reduce response times, and automate repetitive tasks (Dwivedi et al., 2021). They are also used for several purposes other than customer services, such as education, mental health support, and financial management (Ashfaq et al., 2020). It is important to note that chatbots provide 24/7 customer support, contributing to customer satisfaction. Moreover, they can handle multiple conversations concurrently, thereby improving efficiency (Adamopoulou & Moussiades, 2020a). Chatbots such as Siri, Alexa, and Google Assistant are some of the most popular ones (Adamopoulou & Moussiades, 2020b). A collection of notable chatbots is shown in Figure 1 according to their release date.

ChatGPT is a notable chatbot that has emerged in recent times. It is a language model based on Reinforcement Learning from Human Feedback and was created to produce conversational outputs (Thorp, 2023). Over time, ChatGPT has gained widespread recognition and popularity due to its remarkable capacity to generate coherent and realistic responses on a broad range of topics (Lund & Wang, 2023). The implications of this technology for both science and society are noteworthy, as ChatGPT has the potential to impact various industries and fields (van Dis et al., 2023).

This paper aims to analyze the metadata of all the papers indexed in Scopus and WoS that deal with the topics of chatbots and ChatGPT. This study is wide-ranging and appeals to a diverse and inclusive audience, including stakeholders with a vested interest in chatbots and ChatGPT from various fields such as computer programming, education, research, healthcare, and other related areas. As part of this study, we systematically analyze the literature on chatbots with a focus on chatGPT using the "bibliometrix" R package (Aria & Cuccurullo, 2017) as well as VOSviewer software (van Eck & Waltman, 2009), a software tool for visualizing and analyzing bibliographic information. Bibliometric is an academic discipline that utilizes bibliographic data in order to assess and better understand the structure of a particular research area (Keramatfar, Rafiee and Amirkhani, 2022). The paper offers valuable insights regarding the leading countries that focus on publishing research papers on chatbots and ChatGPT and the progression of the topics explored in those publications over time. In addition, we explore other aspects of the analysis, such as sources, authors, and keyword analysis.

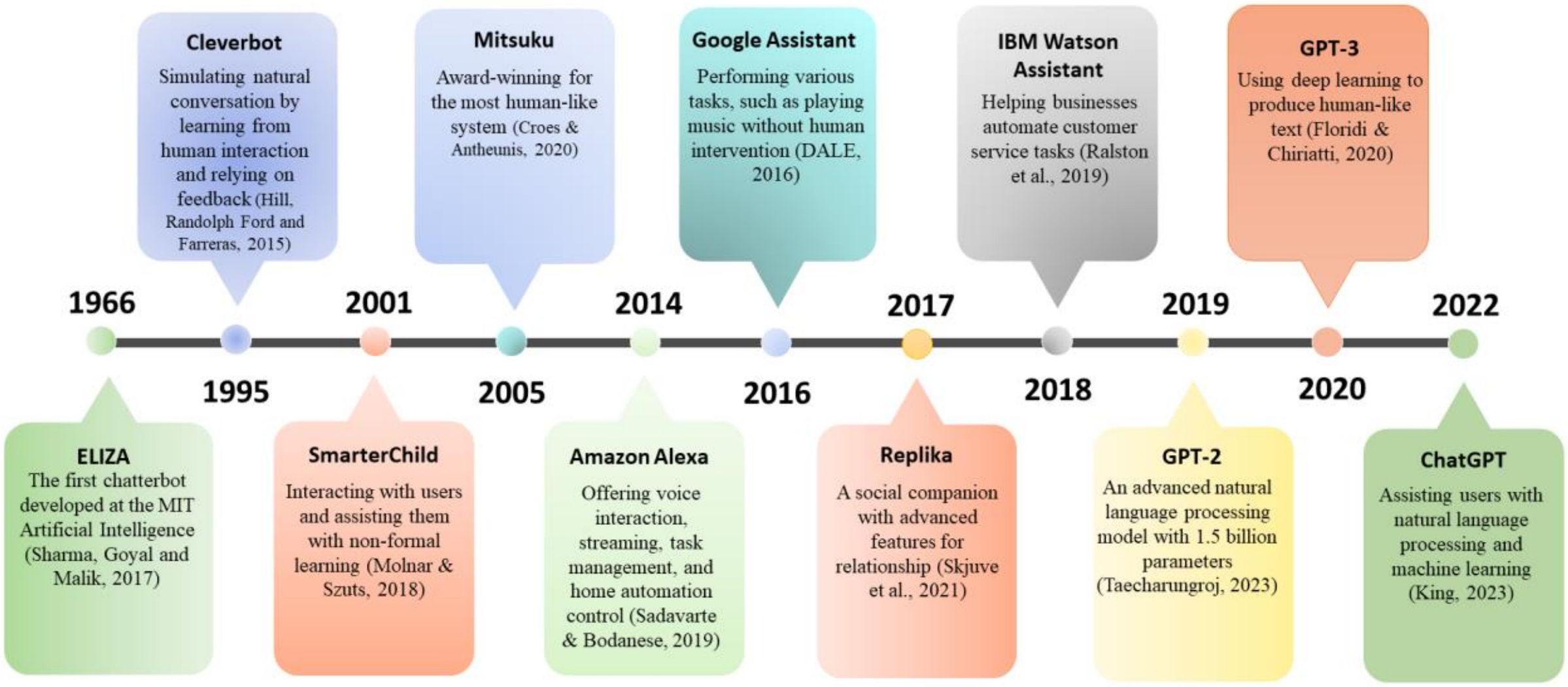

**Fig. 1.** A Summary of well-known chatbots over time

The rest of this paper is organized as follows. Section 2 outlines the methodology implemented in this study. Section 3 presents our findings on bibliometric analysis and reviews of publications. In Section 4, we discuss the critical discoveries of our findings. Finally, Section 5 provides the main conclusions from this study.

## 2. Methodology

In terms of scientific databases, we selected Scopus and Web of Science, since both include the most important journals in computer science, statistics, engineering, and mathematics (Martarelli & Nagano, 2021). In our literature review, we have reviewed a wide range of scholarly sources, including book chapters, articles, reviews, and other relevant publications. In order to ensure that the review is comprehensive and inclusive of all relevant academic and scientific literature, we did not limit our search to specific sources or publication types. Using this approach, we are able to thoroughly evaluate the existing research and knowledge in the field, providing a comprehensive and nuanced understanding of the subject. Overall, the research was conducted in two phases:

I. Initially, a bibliometric analysis was performed on chatbots publications. The analysis included an evaluation of countries, sources, authors, and keywords. The study examined the number of publications from each country, country collaboration, Bradford's Law of sources, author impact, thematic evolution, co-occurrence analyses, and keyword trending topics.

II. Secondly, a bibliometric analysis was conducted on ChatGPT, the latest trend in chatbots. The keywords used in this area were scrutinized using co-occurrence analysis and a word cloud. To gain a more comprehensive understanding, 45 publications on ChatGPT were reviewed. The publications included 15 articles that were evaluated based on the method used, the novelty of the findings, and the conclusions reached. An overview of the methodology can be found in Figure 2.

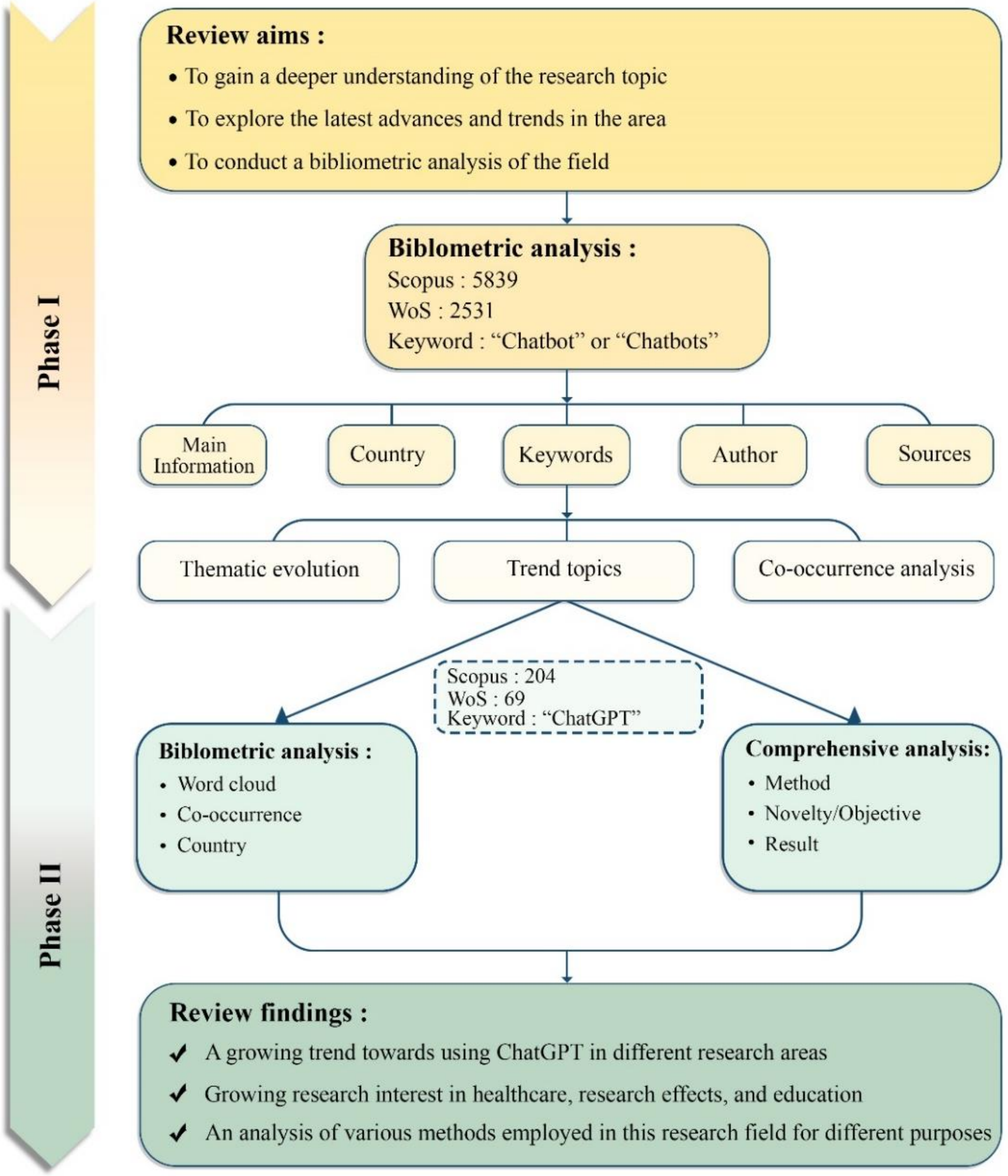

**Fig. 2.** Framework of the methodology utilized in the study

## 3. Results

Bibliometric analysis was carried out using two tools, namely Vosviewer software and bibliometrix (biblioshiny) in the R programming language. The results of the analysis are presented in two separate phases in this section.

## 3.1. Phase I

This section provides a bibliometric analysis of scientific publications on chatbots. The analysis includes information on the countries, sources, authors, and keywords related to the publications. Initially, we provide a general overview of the publications presenting the details of the exported literature.

### 3.1.1. Main Information

As previously mentioned, the study was conducted using the Scopus and WoS databases. The primary information regarding the publications that were exported for the analysis is presented in Table 1.

Table 1: The Main Information about the Exported publication

| Feature | WoS | Scopus |
|---|---|---|
| Timespan | 1998:2023 | 2000:2023 |
| Documents | 2531 | 5839 |
| Sources (Journals, Books, etc) | 1512 | 4813 |
| Authors | 7693 | 14908 |
| Article | 1000 | 1824 |
| Book chapter | 16 | 112 |
| Review | 81 | 179 |
| Annual Growth Rate % | 19.16 | 27.19 |
| Average citations per doc | 7.567 | 8.302 |

Based on the data presented in Table 1, it is evident that the Scopus database exhibits a higher number of publications, authors, book chapters, as well as a higher average citation and annual growth rate, compared to the WoS database. Additionally, we conducted an analysis of the number of common literature published in both databases and discovered that 1946 publications were present in both databases. Thus, the total number of literature analyzed was 6424. It is important to note that the databases were analyzed separately, which could offer more practical insights.

We have calculated the traditional overlap (TO) and the relative overlap (RO) of the Scopus and WoS databases containing studies related to chatbots. For databases X and Y, traditional overlap (TO) is defined as the percentage of overlap between these two databases (Gluck, 1990). This is calculated as:

$$\%TO = 100 \times \frac{|X \cap Y|}{|X \cup Y|}$$

Relative overlap (RO) is used to calculate the overlap or percentage of coverage of a database X, in relation to another database Y and vice versa (Garg, Kumar and Singh, 2020). For any two databases X and Y, the equations for the relative overlap (RO) are given in the following.

$$\%Overlap \in X = 100 \times \frac{|X \cap Y|}{|X|}$$

$$\%Overlap \in B = 100 \times \frac{|X \cap Y|}{|Y|}$$

As seen from Table 1, a total of 2531 and 5839 documents have been obtained from the WoS and Scopus databases, respectively. Of these, 6424 documents have either appeared in WoS or in Scopus and 1946 documents appeared in both the databases. The percentage of TO of the documents from these two databases is 76.887 which is calculated in the following.

$$\%TO = 100 \times \frac{|WoS \cap Scopus|}{|WoS \cup Scopus|} = 100 \times \frac{1946}{6424} = 30.293\%$$

This indicates that between WoS and Scopus databases related to chatbot-focused research, there is a similarity of 30.293%. Similarly, the percentage of RO in the Scopus database is found to be 33.328% whereas the percentage of RO in the WoS database is 76.887%. These are obtained from the following calculations.

$$\%Overlap \in Scopus = 100 \times \frac{|WoS \cap Scopus|}{|Scopus|} = 100 \times \frac{1946}{5839} = 33.328\%$$

$$\%Overlap \in WoS = 100 \times \frac{|WoS \cap Scopus|}{|WoS|} = 100 \times \frac{1946}{2531} = 76.887\%$$

These results indicate that WoS covers 33.328% of the documents in Scopus whereas Scopus covers 76.887% of the documents in WoS.

#### 3.1.2. Country Analysis

As of the end of March 2023, the WoS database contained 2531 documents on chatbots from 92 countries/territories. Scopus, on the other hand, contained 5839 chatbot-related documents from 116 countries as of the same date. These findings suggest that chatbot research is a rapidly growing field, with a significant global presence. The United States is the clear leader in this field, with the most publications on both databases (897 and 1425 in Scopus and WoS, respectively). This country has the most collaboration with other countries in WoS. The United Kingdom, Germany, and Italy are also major players, forming the largest collaborative cluster on Scopus.

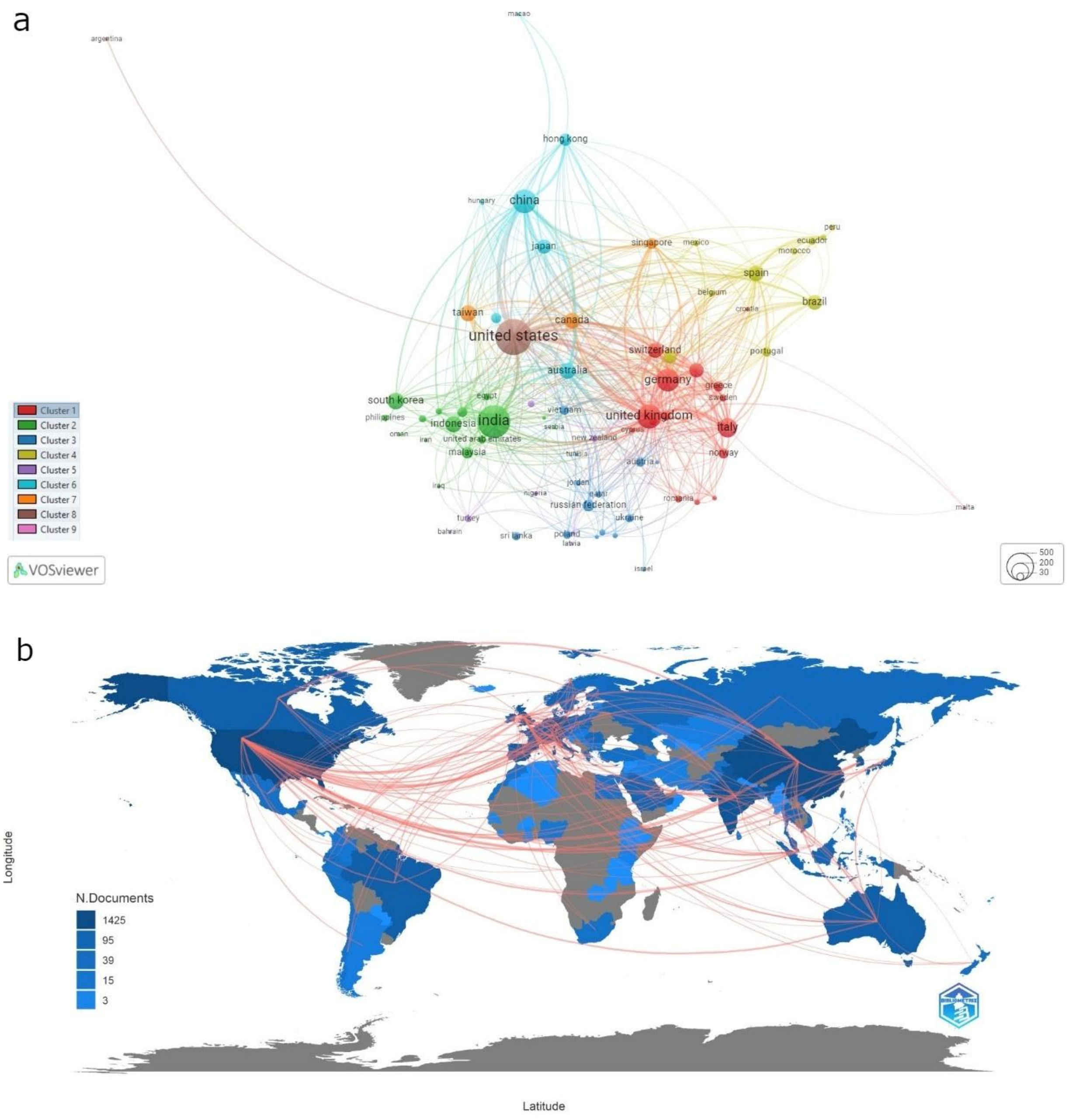

**Fig. 3.** Collaboration Network of Countries on Scopus (a) and Geographical Distribution of Publications in Countries on WoS (b

### 3.1.3. Sources (Bradford's Law)

According to Bradford's (1934) law, the number of papers on a subject is distributed (scattered) according to a mathematical function, so that the number of journals/information sources must increase to accommodate an increase in papers on the subject. Groups of journals producing nearly equal numbers of articles are roughly proportional to $1{:}\,n{:}\,n^2$ ..., where $n$ is the Bradford multiplier (Sangam, 2015). The top third (Zone 1 or core) of any issue or subject area represents journals that are the most frequently cited in the literature of the subject, and that are, therefore, likely to be of the greatest interest to researchers. The middle third (Zone 2) consists of journals that have been cited on average, whereas the bottom third (Zone 3 or tail) consists of journals that are rarely cited and considered marginally important (Potter, 2010). Bradford analyses offer a range of potential applications, including the selection and deselection of journals, the identification of the core set of relevant publications, and the evaluation of data collection (Nisonger, 2008). Also, it can be employed to make predictions regarding the extent of relevant information that may be absent from an incomplete search (Nash-Stewart et al., 2012).

### 3.1.3.1. Verbal Formulation

In order to test the verbal formulation of Bradford's Law, the cited literature is organized in descending order of the number of published literature. The rank number of each source, along with its corresponding number of documents and cumulative documents are provided in Table 2. 1914 scientific documents were published in 32 sources, followed by 1972 articles in 525 journals. The remaining were appeared in 1732 sources, meaning each journal category embraced one-third of the total literature. In the current setting, each Zone's relationship in Bradford's distribution is $32\colon 525\colon 1732$, where the value multiplier is 54.13.
Based on Table 2, the calculation for Bradford's Law is as follows:

$$32\colon 32 * 54.13\colon 32 * (54.13)^2 :: 1\colon n\colon n^2$$
$$32\colon 1732.16\colon 93761.82$$

So, the percentage error is given as:

$$\frac{95525.98 - 2289}{2289} \times 100 = 40.73 \times 10^2$$

Table 2: Distribution of journals and citations across the Bradford Zone

| Zones | Number of Sources | Percentage of Sources (%) | Number of Documents | Cumulative No. of Documents | Bradford Multiplier |
|---|---|---|---|---|---|
| 1 | 32 | 1.40 | 1914 | 1914 | 1 |
| 2 | 525 | 22.94 | 1972 | 3886 | 16.41 |
| 3 | 1732 | 75.66 | 1916 | 5802 | 54.13 |
| **Total** | **2289** | **100** | **5802** | | |

As a result, the data cannot be fitted into Bradford's Law due to a high percentage error. Therefore, it is necessary to assess the adherence of the data to Bradford's Law of Scattering using the Leimkuhler Model (LEIMKUHLER, 1980), which has been employed in various previous studies (Bhargav, Kishore and Doraswamy, 2020; Qiu et al., 2017).

### 3.1.3.2. Application of Leimkuhler Model

The Leimkuhler Model, which employs Bradford's distribution, first identifies the core journals with distinct references in the first Zone. This helps to determine Bradford's multiplier (k). As a size-frequency measure, the Leimkuhler Model then uses multiples of Bradford's multipliers to identify the number of journals in the subsequent Zones. Egghe's mathematical technique (Egghe, 1990) is used to calculate the multiplier (k) for the Leimkuhler distribution.

Leimkuhler's Model is defined in accordance with Bradford's verbal formulation as follows (Borgohain et al., 2021).

$$r_0 = \frac{T(k-1)}{(k^p-1)} \tag{1}$$

where $T$ represents the total number of journals, $k$ indicates the Bradford's multiplier, and $p$ is the number of Zones. In this study, the articles are divided into three Zones so that the mathematical expression that implements Bradford's law can be expressed as follows.

$$k = (e^y \times y_m)^{\frac{1}{p}}$$

where $y_m$ is the number of items in the most productive source. Therefore, Bradford's multiplier $k$ is calculated as,

$$y_m = 421$$

$$p = 3$$

$$k = (1.781 \times 421)^{\frac{1}{3}} = 9.08$$

Also,

$$y_0 = \frac{A}{p} = \frac{5802}{3} = 1934$$

where $A$ is the total number of articles. Now, $r_0$ is calculated in the following.

$$r_0 = \frac{T(k-1)}{(k^p-1)} = \frac{2289(9.08-1)}{(9.08^3-1)} = 24.97$$

Using the value of $log(k)$ we can obtain the following.

$$a = \frac{y_0}{log(k)} = \frac{1934}{log(9.08)} = 2018.61$$

$$b = \frac{k-1}{r_0} = \frac{9.08-1}{24.97} = 0.32$$

Thus,

$$r_1 = r_0 \times k = 24.97 \times 9.08 \simeq 225$$

$$r_2 = r_0 \times k^2 = 24.97 \times 9.08^2 \simeq 2025$$

Therefore, the number of journals in the nucleus, $r_0$ is 25 and the mean value of the Bradford multiplier, $k$ is 9.08. Hence, Bradford's distribution from the Leimkuhler Model can be expressed as:

$$r_0 : r_0 * k : r_0 * k^2$$

Therefore, we get,

$$25 : 25 * 9.08 : 25 * (9.08)^2 :: 1 : n : n^2$$
$$25 : 227 : 2058$$

Finally, the percentage error is calculated as,

$$\frac{2310 - 2289}{2289} \times 100 = 0.9$$

Table 3: Leimkuhler Model of Bradford's Distribution in 3 Zones

| Zones | Number of Sources | Percentage of Sources (%) | Number of Documents | Cumulative No. of Documents | Bradford Multiplier |
|---|---|---|---|---|---|
| 1 | 25 | 1.08 | 1807 | 1807 | 1 |
| 2 | 227 | 9.83 | 1365 | 3172 | 9.08 |
| 3 | 2058 | 89.09 | 2630 | 5802 | 82.45 |
| **Total** | **2310** | **100** | **5802** | | |

The calculation showed a very small percentage error of 0.9%, which can be considered negligible. The results of the analysis using the Leimkuhler Model to assess Bradford's scattering are presented in Table 3. The table indicates that the number of journals representing documents in all three Zones increases by a factor of approximately 9. The zonal analysis of the data showed that Zone 1 comprises 25 sources and 1807 documents, Zone 2 consists of 227 sources and 1365 documents, and Zone 3 consists of 2058 sources and 2630 documents.

### 3.1.4. Authors

This section ranks the top 10 authors in chatbot research based on their total citations. The authors' number of publications (N), h-index, g-index, m-index, and year of first publication (YS) in the area are also included. Then, the collaboration between authors in both databases is shown.

Table 4: Top 10 Authors' impacts based on Total Citation

| Scopus | | | | | | | WOS | | | | | | |
|---|---|---|---|---|---|---|---|---|---|---|---|---|---|
| **Author** | **TC** | **h_index** | **g_index** | **m_index** | **N** | **YS** | **Author** | **TC** | **h_index** | **g_index** | **m_index** | **N** | **YS** |
| Gao J | **1344** | 6 | 8 | 0.75 | 8 | 2016 | Folstad A | **380** | **7** | **13** | **0.395** | **13** | 2017 |
| Zhou M | 1178 | 9 | 13 | 0.529 | 13 | 2007 | Sundar SS | 375 | 3 | 4 | 0.375 | 4 | **2016** |
| Littman ML | 1170 | 1 | 1 | 0.048 | 1 | **2003** | Li D | 345 | 2 | 3 | 0.333 | 3 | 2018 |
| Turney PD | 1170 | 1 | 1 | 0.048 | 1 | **2003** | Singh D | 312 | 2 | 3 | 0.286 | 3 | 2017 |
| Galley M | 1049 | 4 | 4 | 0.5 | 4 | 2016 | Batra D | 311 | 2 | 2 | 0.286 | 2 | 2017 |

| | | | | | | | | | | | | |
|---|---|---|---|---|---|---|---|---|---|---|---|---|
| Følstad A | 1042 | **13** | **25** | **1.857** | **25** | 2017 | Das A | 311 | 2 | 2 | 0.286 | 2 | 2017 |
| Wu W | 979 | 11 | 18 | 1.375 | 18 | 2016 | Gupta K | 311 | 2 | 2 | 0.286 | 2 | 2017 |
| Wu Y | 896 | 10 | 16 | 0.625 | 16 | 2008 | Kottur S | 311 | 2 | 2 | 0.286 | 2 | 2017 |
| Xing C | 803 | 6 | 6 | 0.75 | 6 | 2016 | Moura JMF | 311 | 2 | 2 | 0.286 | 2 | 2017 |
| Brandtzaeg PB | 771 | 9 | 12 | 1.286 | 12 | 2017 | Parikh D | 311 | 2 | 2 | 0.286 | 2 | 2017 |

According to the table 4, Folstad A has the highest number of citations in the WOS database and ranks as the sixth-most cited author in the Scopus database. This is notable because Folstad A is the only author who appears on both databases' list of the top 10 most cited authors. Additionally, Folstad A has the highest h-index, g-index, and m-index in both databases, further highlighting his significant impact on the field of study. These bibliometric indicators suggest that Folstad A has made a notable contribution to the field and is highly regarded by other researchers in the community. Also, the data reveals that Gao J is the most cited author in Scopus, with a total citation count of 1344. Despite having only eight publications, Gao J has an h-index of 6, which indicates that at least 6 of his papers have been cited six or more times. In Figure 4, network analysis was conducted to identify key authors and their collaboration networks in both databases. The analysis revealed 15 clusters comprising nearly 350 and 250 authors (for Scopus and WOS, respectively) included in the study.

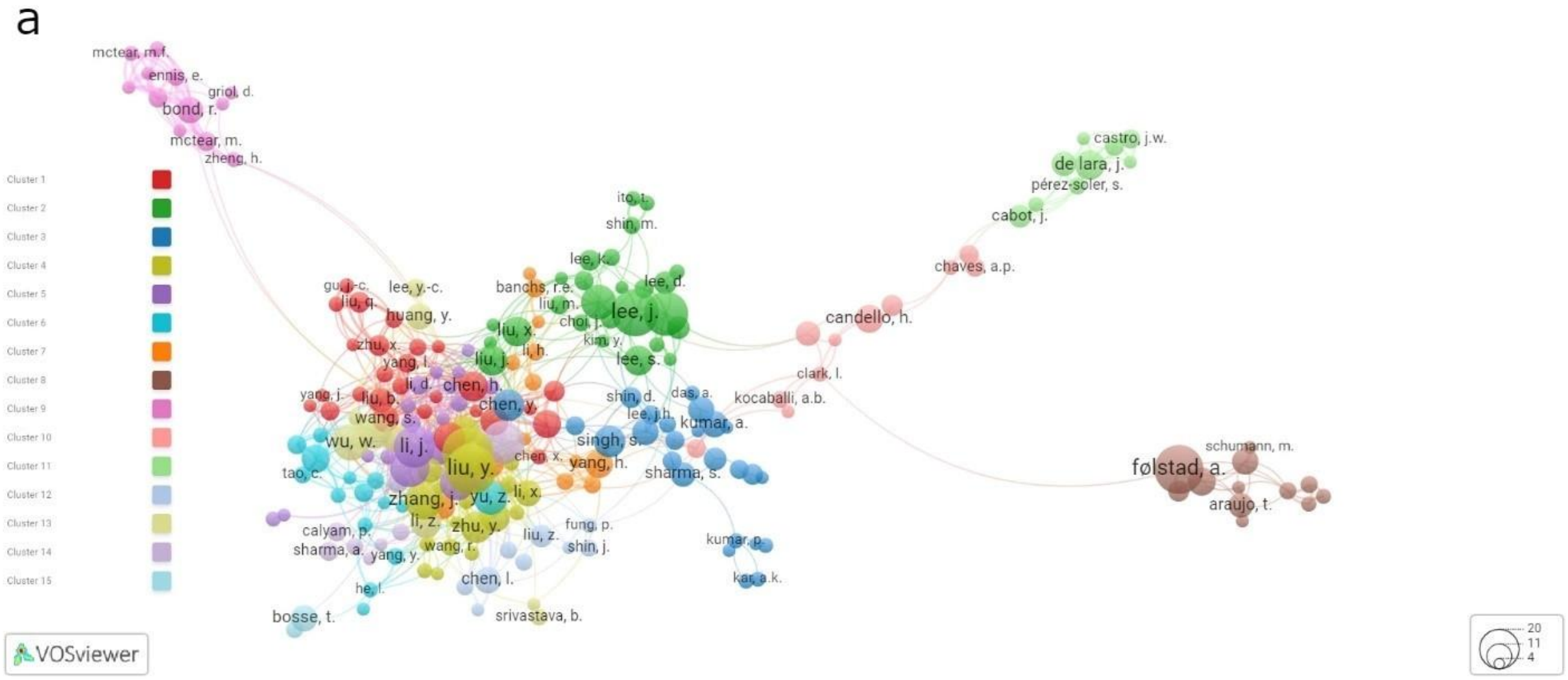

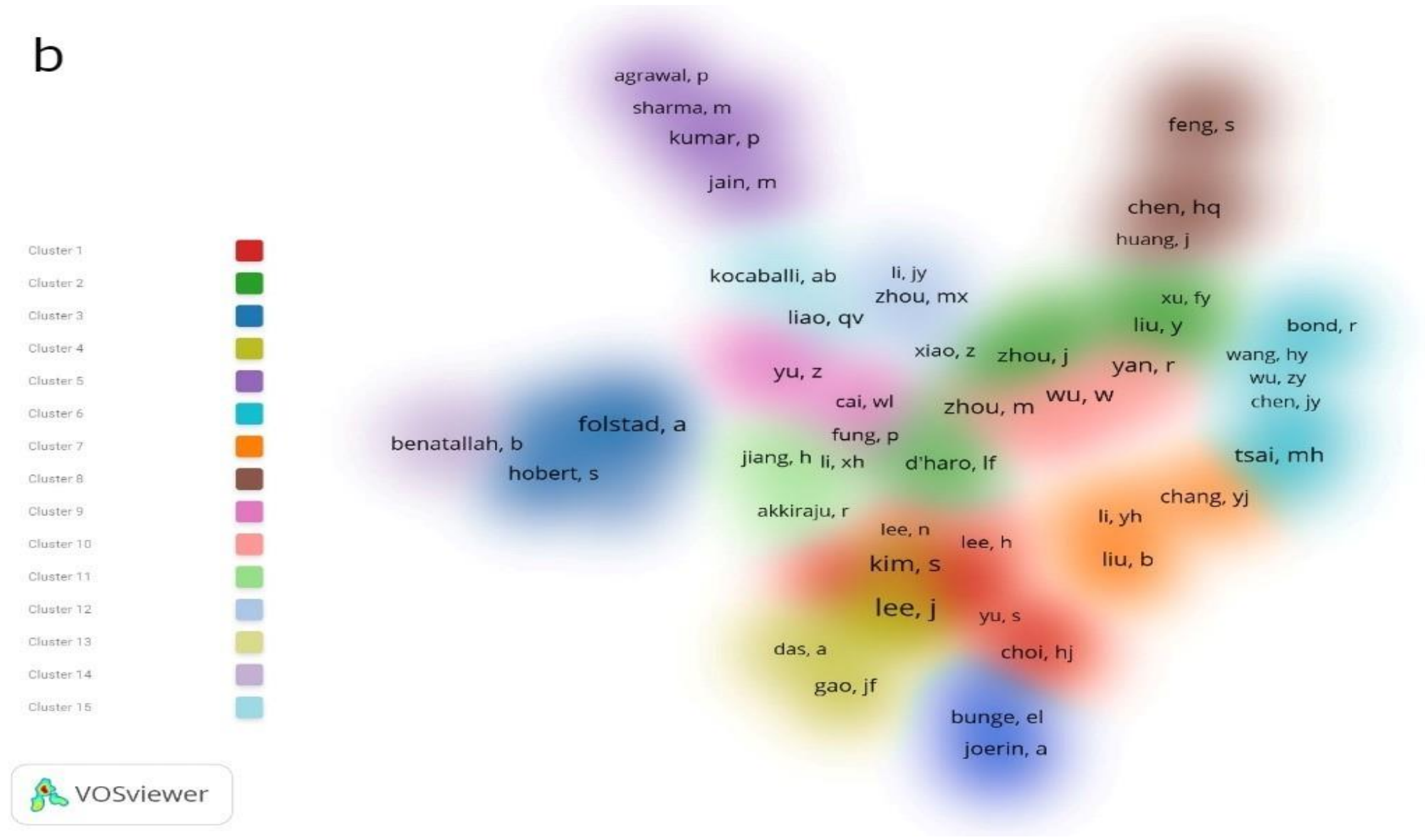

**Fig. 4.** Collaboration Network of Authors on Scopus (a) and WoS (b)

Figure 4 analyzes collaboration patterns between authors in Scopus and WoS databases using the association strength method based on document weights. The 15 clusters are ordered according to the number of authors in each group. The collaboration between specific authors, such as Lee J, Kim S, and Lee D, contributed significantly (Clusters 2 and 4 in Scopus and WoS, respectively). These clusters included authors ranging from 2 to 27 in Scopus and 5 to 16 in WOS. Overall, the findings suggest that collaboration patterns between authors differ across databases. Also, author Folstad A's impact was confirmed by this Figure.

### 3.1.5. Keywords

The evolution and relationships between research topics are presented in Figure 5, divided into three analysis periods. The characteristics of the lines indicate the quality of relationships between keywords. The study divided the twenty-three-year and twenty-five-year periods for Scopus and WoS, respectively, into four and five periods, namely [2000-2003], [2004-2014], [2015-2017], [2018-2023] for Scopus, and [1998-2006], [2007-2014], [2015-2017], [2018-2020], and [2020-2023] for publications in WoS.

Figure 5 illustrates the strong interconnectivity of research topics in the field of study. The figure reveals that research topics have links to different research areas. In the 2015-2017 period, deep learning and machine learning emerged as essential topics in the field. Moreover, recent research has shown that topics such as natural language processing (NLP) and mental health have gained significant interest in the field of study. Figure 5b highlights the recent trust and task analysis research areas, indicating their growing importance in the field. These findings demonstrate the importance of emerging research areas and provide valuable information about research topics structure and evolution. Next, Figure 6 displays the co-occurrence network of keywords in the relevant field of study. The clustering algorithm employed in this analysis was walktrap, with a selection of 50 keywords for clustering, and normalization was conducted by association measures.

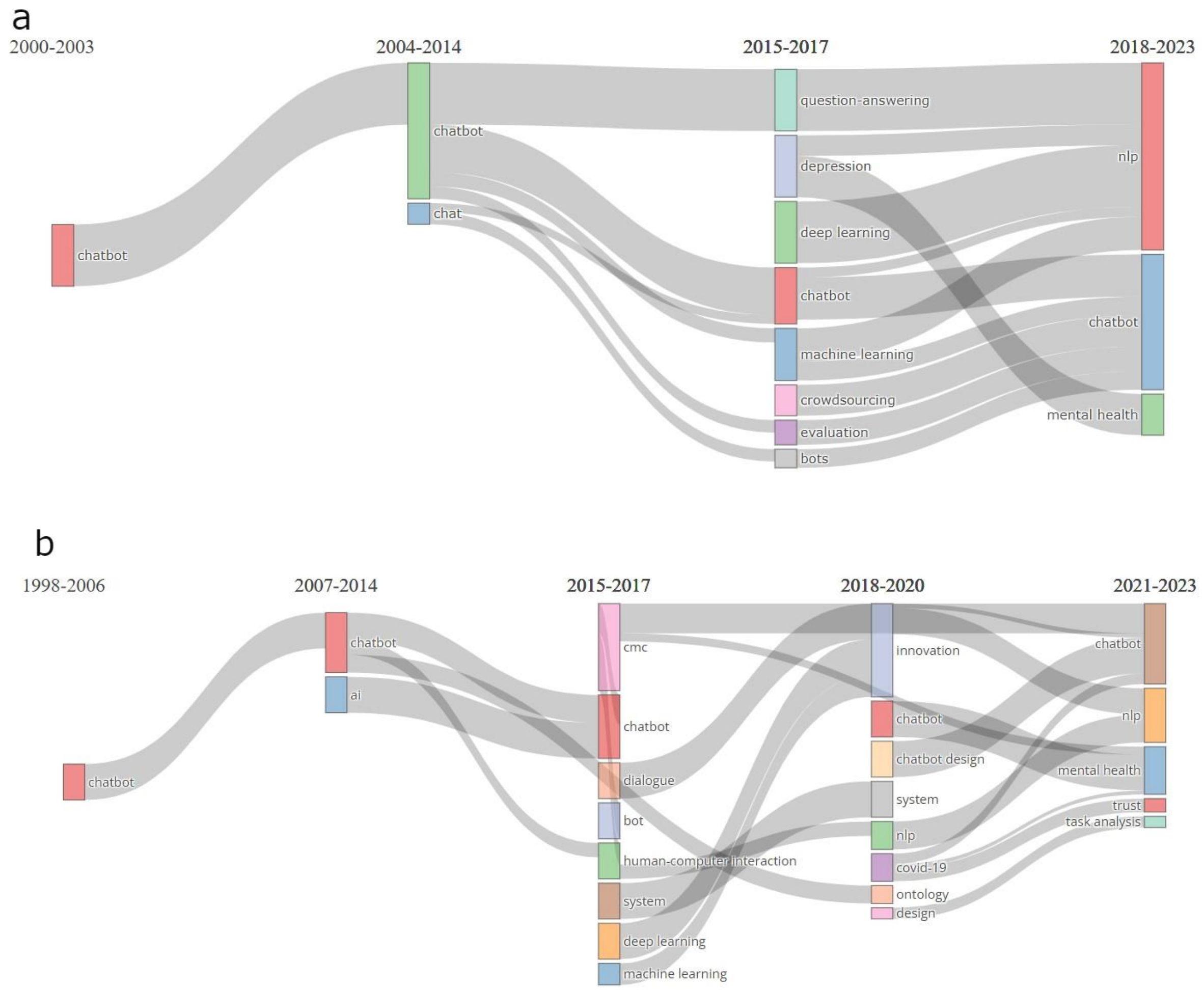

**Fig. 5.** Thematic Evolution of Keywords on Scopus (a) and WoS (b)

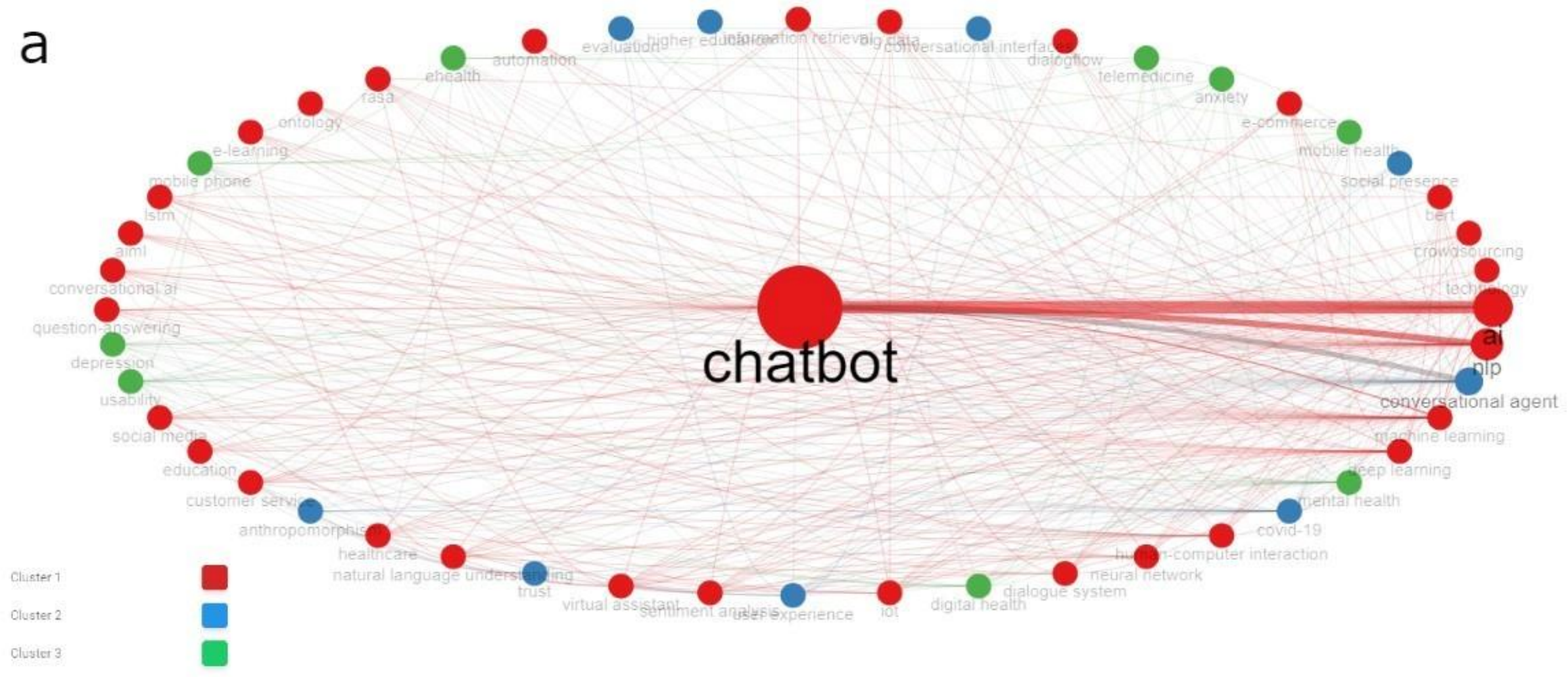

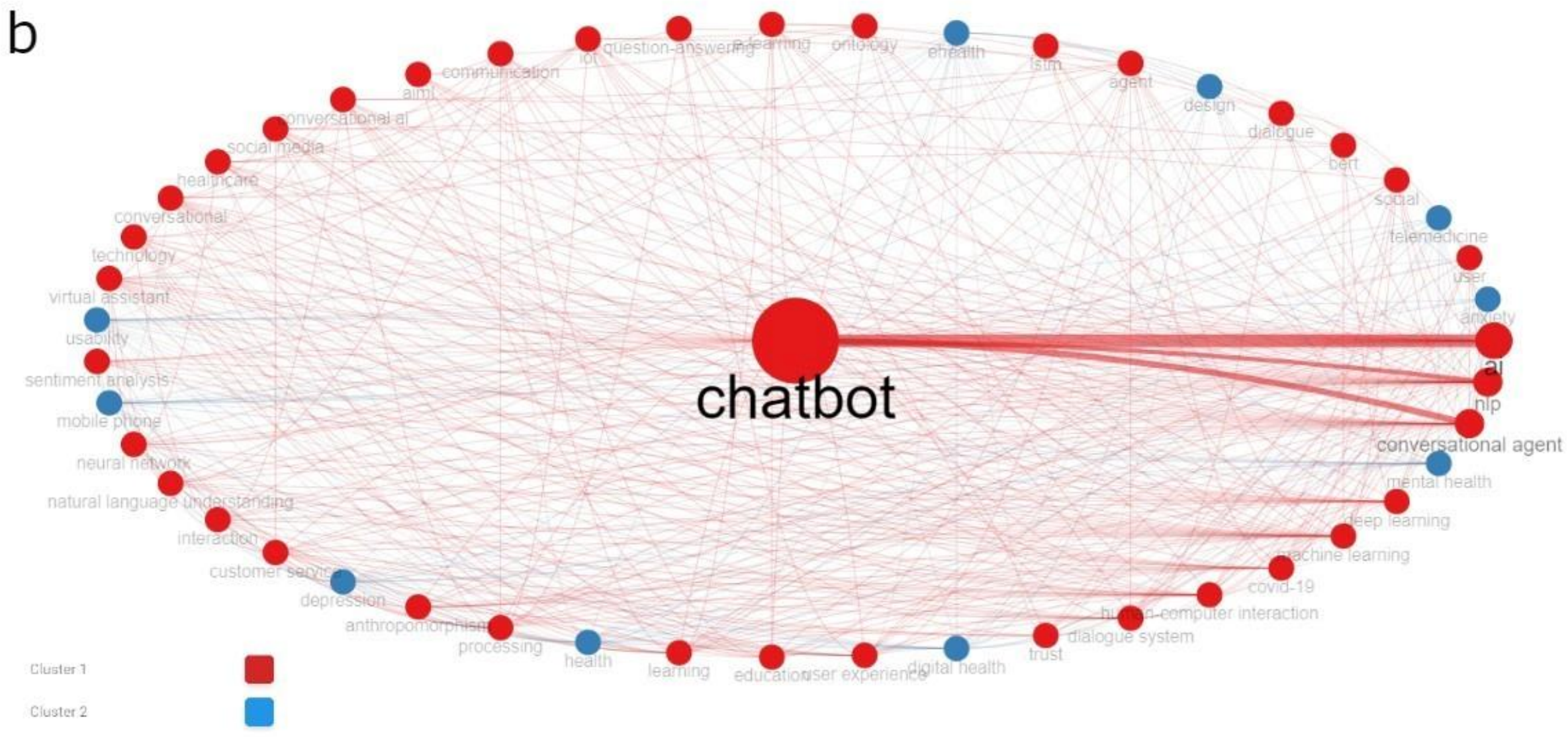

**Fig. 6.** Co-occurrence Network of Keywords on Scopus (a) and WoS (b)

Based on Figure 6, three distinct clusters of keywords based on Scopus publication were identified. These clusters were labelled "artificial intelligence and related technologies", "design and evaluation of conversational agents", and "digital technologies and mental health". Alternatively, the keywords extracted from the WoS can be divided into two categories: "AI-based chatbots and conversational agents" and "digital technologies and mental health". It is worth noting that clusters 1 and 2 from Scopus were merged into a single cluster in WoS. These findings suggest that a significant amount of research is being conducted in these areas and that there is potential for further interdisciplinary research at the intersection of artificial intelligence, digital technologies, and health. Tables 5 and 6 present the top 5 keywords within each cluster for Scopus and WOS databases, respectively. This provides valuable insights into the main topics and themes being researched in each cluster.

Table 5: Top 5 keywords within each cluster for Scopus

| Keyword | Cluster | Betweenness | Closeness |
|---|---|---|---|
| chatbot | 1 | 663.1045 | 0.020833 |
| ai | 1 | 65.05425 | 0.020833 |
| nlp | 1 | 25.86692 | 0.017544 |
| machine learning | 1 | 8.302174 | 0.016949 |
| deep learning | 1 | 2.358797 | 0.014085 |
| human-computer interaction | 1 | 0.621563 | 0.013158 |
| conversational agent | 2 | 27.84586 | 0.018868 |
| covid-19 | 2 | 2.459913 | 0.014925 |
| user experience | 2 | 0.324901 | 0.012987 |
| trust | 2 | 0.066949 | 0.011628 |
| anthropomorphism | 2 | 0.105056 | 0.011494 |
| mental health | 3 | 4.926437 | 0.014286 |
| digital health | 3 | 1.037543 | 0.013158 |
| usability | 3 | 0.295981 | 0.012821 |
| depression | 3 | 0.209681 | 0.012346 |
| mobile phone | 3 | 0.105715 | 0.012195 |

Table 6: Top 5 keywords within each cluster for WOS

| Keyword | Cluster | Betweenness | Closeness |
|---|---|---|---|
| chatbot | 1 | 572.2521 | 0.021277 |
| ai | 1 | 38.31512 | 0.02 |
| nlp | 1 | 22.55187 | 0.019231 |
| conversational agent | 1 | 20.88957 | 0.018519 |
| deep learning | 1 | 2.363443 | 0.014286 |
| mental health | 2 | 2.804667 | 0.014085 |
| digital health | 2 | 1.222346 | 0.014493 |
| health | 2 | 0.489457 | 0.014286 |
| depression | 2 | 0.582268 | 0.013514 |
| mobile phone | 2 | 0.587108 | 0.013889 |

As we can see in tables 5 and 6, the betweenness and closeness of each keyword are shown. In a network, betweenness refers to how often a node is located at the shortest distance between two other nodes (keywords). Nodes with high betweenness play a crucial role in connecting networks and are pivotal to the overall operation of the network. Furthermore, the closeness indexes reflect the average length of the shortest paths in a network, and the closer the node is to the closest node in the network, the more powerful it tends to be in the network and the more accessible it is to other nodes (Makkizadeh & Sa'adat, 2017). Chatbot has the highest betweenness and closeness in both databases, followed by AI. Conversational agents have also a high betweenness in cluster 2 of Scopus but is in cluster 1 in WoS, which suggests that it is a more important keyword in Scopus than in WoS. This could be because Scopus covers a wider range of academic disciplines than WoS. In the next part of our analysis, we studied the most recent popular topics in the field of chatbots to help researchers and developers to stay up-to-date on the latest developments in the field. It can also help them to identify new opportunities and to avoid potential pitfalls.

According to Figure 7, ChatGPT appears to be the latest trend in the chatbot field, as observed in both databases. We applied a minimum frequency of 5 for keywords and set the number of keywords for each year as 3 in this analysis. As such, we conducted a more detailed investigation into this subject in the second phase of this paper, which features a comprehensive review of published literature and bibliometric analysis.

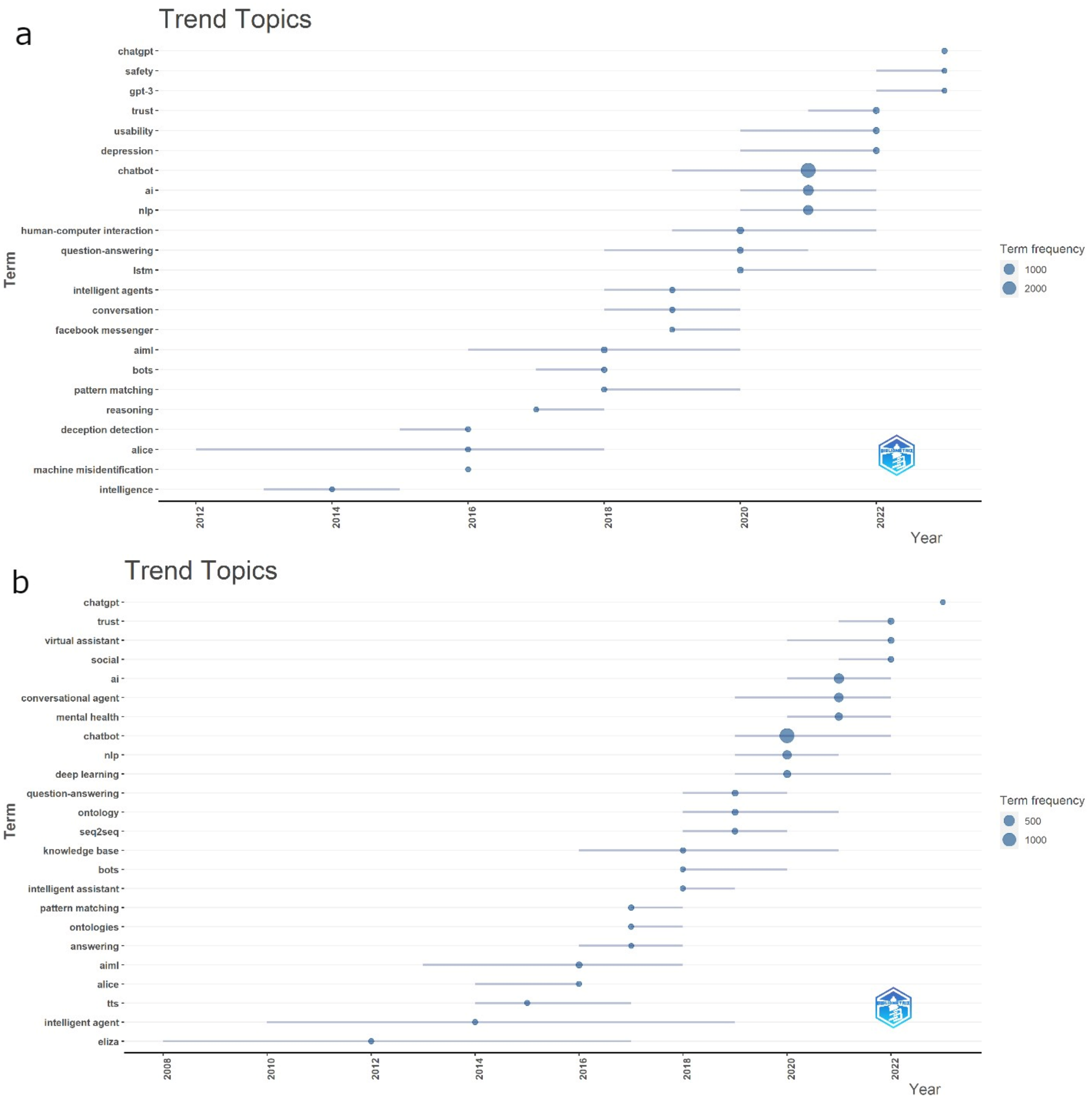

**Fig. 7.** Trend Topics based on the Keywords of the Study Area on Scopus (a) and WOS (b)

## 3.2. Phase II

During this phase, we comprehensively analyzed ChatGPT scientific publications available in Scopus and WoS databases. As of April 2023, 204 publications from Scopus and 69 publications from WoS were considered. Out of these, 62 records were found to be present in both databases. Therefore, for our analysis, we considered 211 scientific publications in total. Our analysis involved an initial bibliometric assessment, followed by a detailed examination of 45 selected publications.

### 3.2.1. Bibliometric Analysis

In this section, 211 publications on ChatGPT are analyzed, 62 of which are classified as articles. The average citation rate is 0.98 per document, and 644 authors contributed to the publications. The publications include 334 keywords, and 78 publications are single-author documents. The documents examined in this study cited approximately 5000 sources. First, the distribution of the publications concerning single and multi-country contributions is presented in Figure 8.

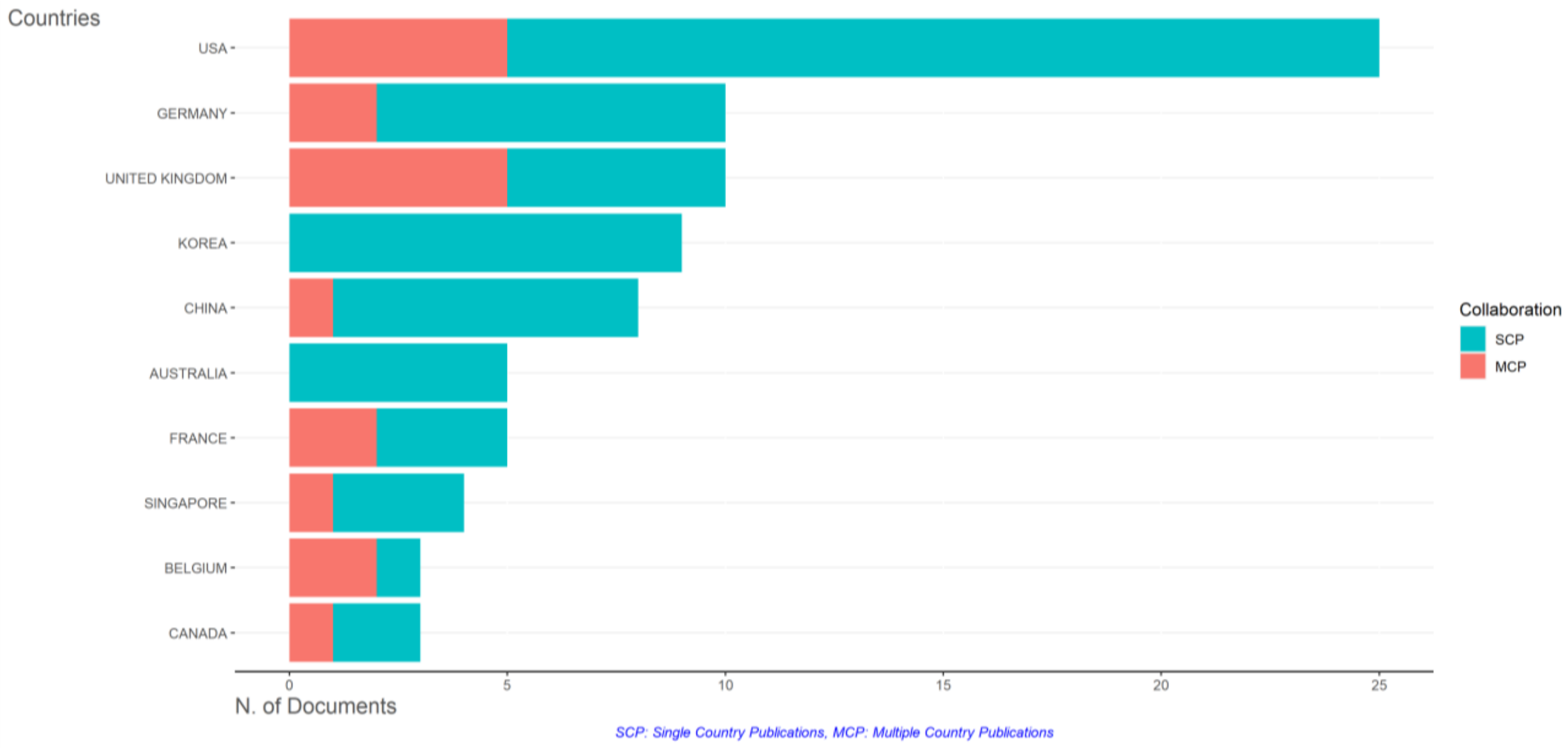

**Fig. 8**. Top 10 countries with highest number of documents

According to Figure 8, the United States had the highest number of documents, with 25 publications. Additionally, the US and UK had the highest number of publications from a single country, with five documents each. To examine the citation patterns of documents based on their country of origin, Figure 9 presents the total number of citations received by each country.

As shown in Figure 9, the United Kingdom had the most cited documents (17), with an average citation rate of 1.7 per article. This rate is higher than that of the United States, which had the highest number of publications (20), at 0.64. Ireland had the highest average citation rate of 3. The next step is to analyze the keywords in this field, and Figure 8 displays a word cloud of the keywords. The word cloud is a visual representation of the frequency of each word in the text. The larger the word, the more frequently it appears in the text.

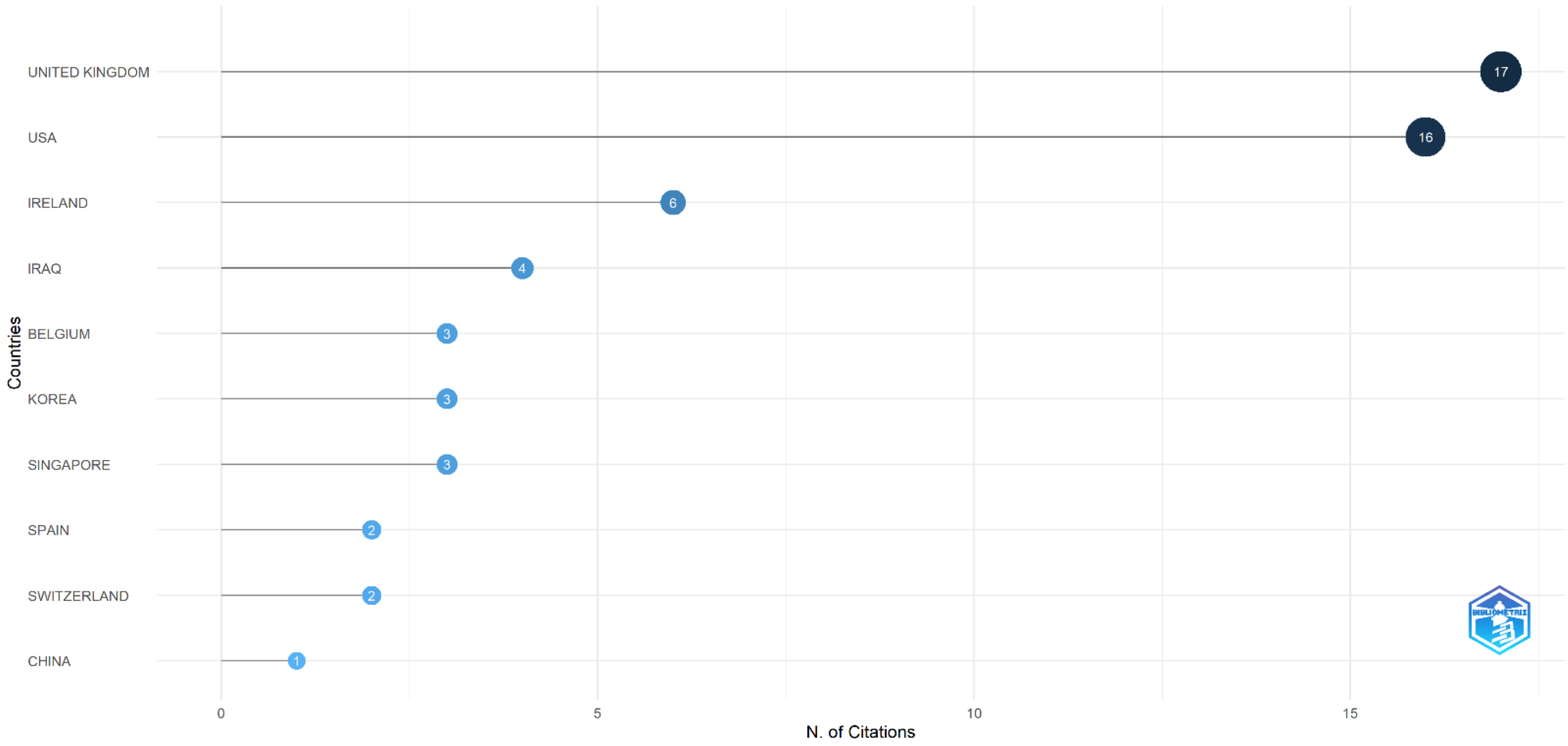

**Fig. 9.** Top 10 countries with highest number of citations

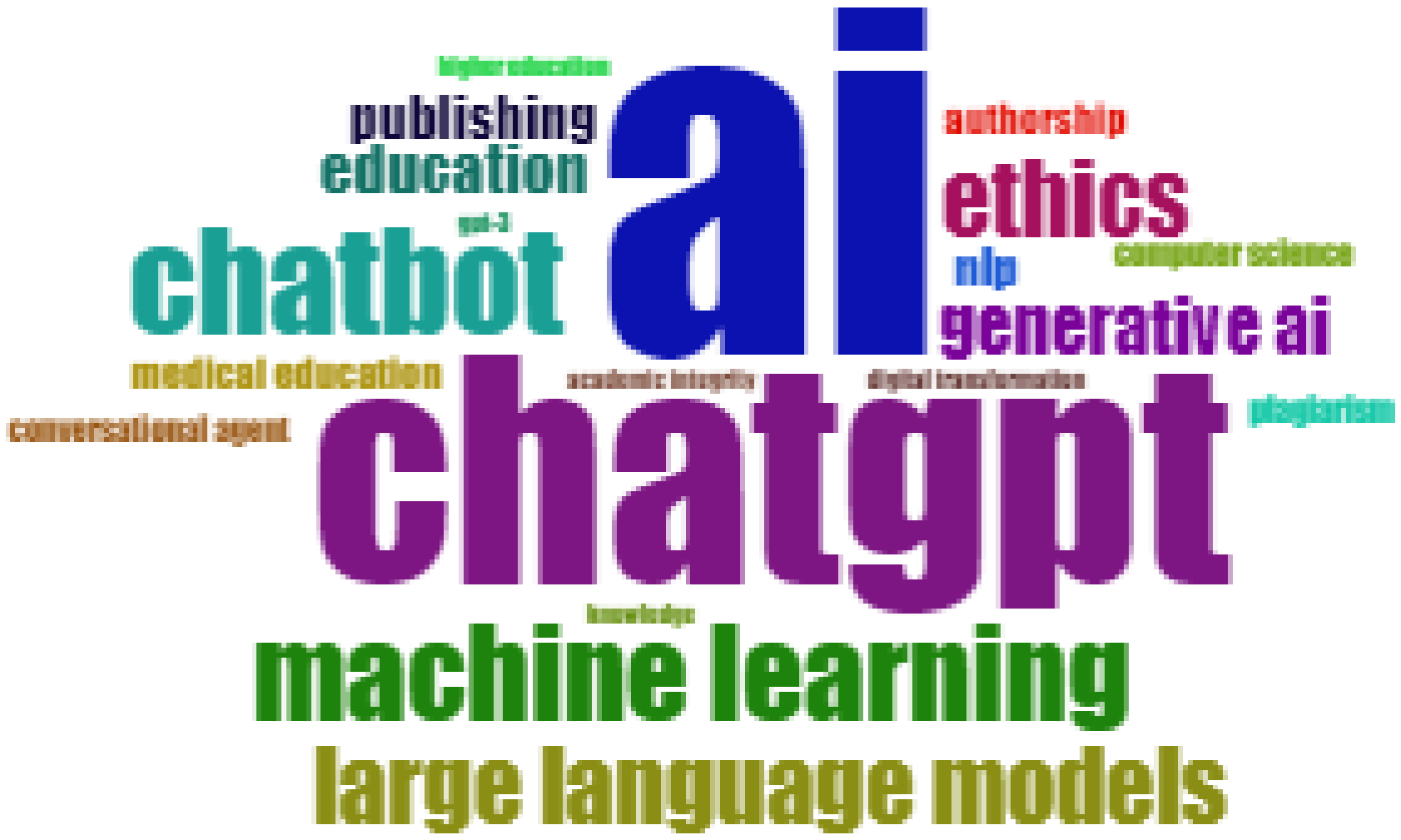

**Fig. 10.** Word cloud of the keywords of literature on ChatGPT

The word cloud in Figure 8 depicts the most frequently used keywords in the field. The most prominent word is "artificial intelligence," followed by "ChatGPT" and "chatbot." These words were used as keywords 55, 36, and 17 times, respectively. "Publishing," "plagiarism," and "authorship" indicate that research ethics is one of the essential concerns in this area. To further analyze the keywords, Figure 11 shows the co-occurrence of keywords.

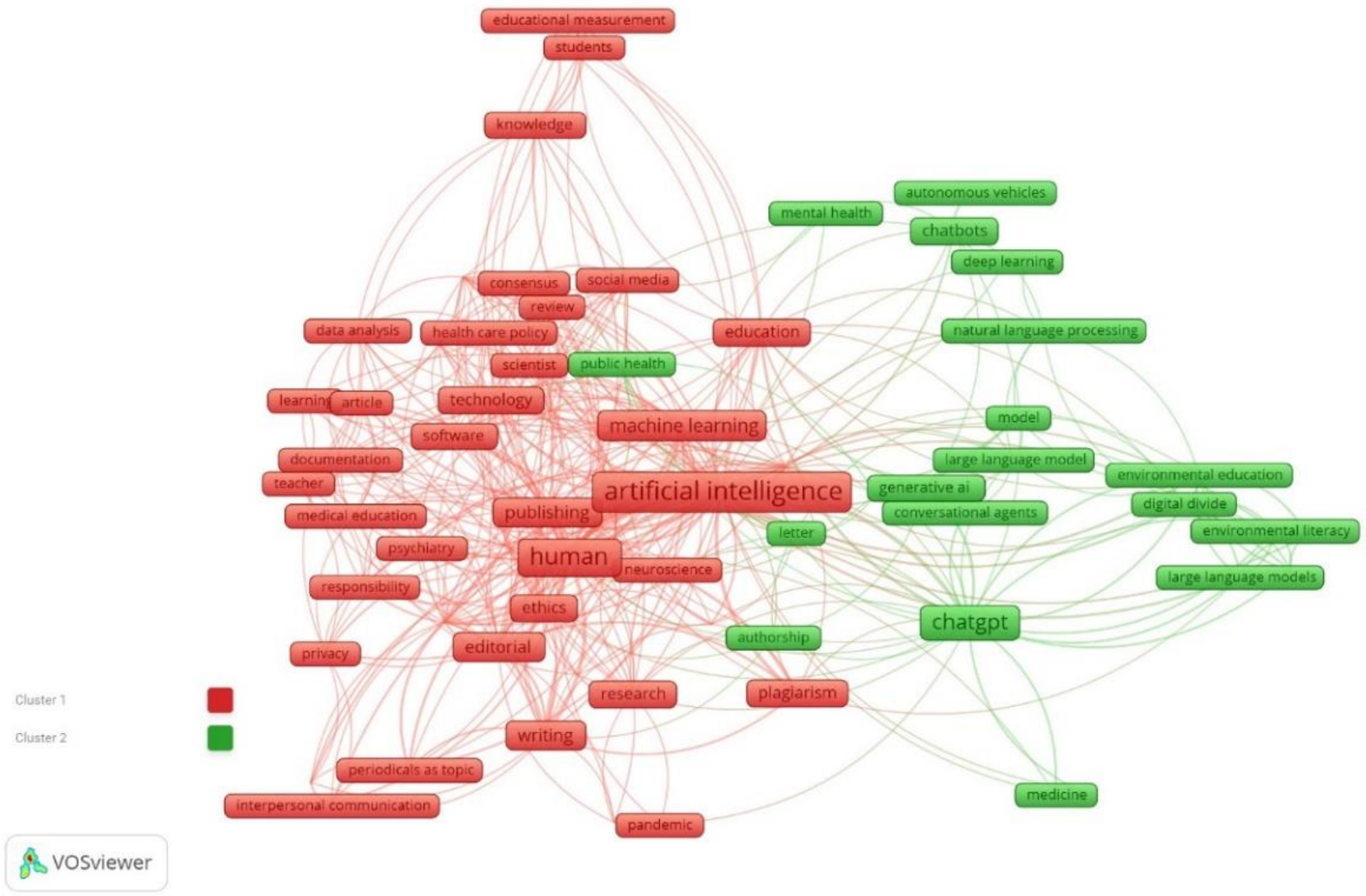

**Fig. 11.** Co-occurrence of the keywords of literature on ChatGPT

Based on Figure 11, two clusters focus on distinct aspects of language models, generative AI, and conversational agents. Specifically, Cluster 1 emphasizes the employment of these technologies in education and writing, whereas Cluster 2 prioritizes their utilization in medicine and healthcare. It is essential to acknowledge, however, that these two clusters are not mutually exclusive, as language models and conversational agents could be employed in both fields. Moreover, the shared keyword "artificial intelligence" suggests that the two clusters are linked by their mutual concern for applying artificial intelligence in different domains. The following section will provide a more detailed analysis of some papers that fall into these two clusters.

### 3.2.2. Comprehensive Review

In this section, 45 published literatures, including 15 papers on ChatGPT, have been fully analyzed. First, table 7 summarizes the 15 articles chosen based on their citation, number of reads, and importance of their domain. Then, review of publications on other sources is provided.

#### 3.2.2.1. Articles

Table 7: A Summary of 15 Articles on ChatGPT

| Authors | Country | Field of study | Novelty/Objective | Method | Results |
|---|---|---|---|---|---|
| (Nachshon et al., 2023) | ISR | Medicine | Discussed a case of 100% third-degree burns and investigated ChatGPT's efficiency in writing a case report | Clinical assessment using queries | Emphasizing the inability of ChatGPT to synthesize the case's emotional or moral aspects by presenting only facts |
| (Rockwell et al., 2023) | USA | Medicine | Generated a manuscript with the help of ChatGPT which discussed the diagnosis and treatment of a patient with a rare fungal infection. | Treatment with systemic antifungals using detailed queries | Generating an article using ChatGPT which found that Histoplasmosis presentation can be highly variable, and diagnosis can be challenging |
| (Macdonald et al., 2023) | UK | Medicine | Evaluated ChatGPT's ability to accelerate the drafting of papers, determining vaccine effectiveness, and generating codes to perform the survival analysis and calculate hazard ratios. | A description of the simulated dataset was provided to ChatGPT and codes were generated through query and error feedback to ChatGPT allowing it to self-correct. | Explaining the prospects of ChatGPT as a great help to researchers in generating code, interpretation of data, and drafting papers |
| (Salvagno, Taccone and Gerli, 2023) | Belgium, Italy | Medicine | Explored the capabilities of ChatGPT in creating initial drafts, organizing the manuscript, and checking of errors in scientific articles in the field of critical care medicine. | Literature review, case studies and future applications | Elucidating the need for human intervention in making significant decisions with consideration of ethical aspects in scientific writing despite ChatGPT's scientific writing capabilities |
| (Mann, 2023) | USA | Medicine | Discussed the role of AI in translational medicine | Conducting interviews with ChatGPT including analysis of large datasets | Importance of addressing the issues related to data dependency, complexity of biological systems, and ethical implications despite having many advantages in translational medicine including efficient and effective analysis of large data |
| (Charrois-Durand et al., 2023) | Canada | Medicine | Examined the ability of ChatGPT to prepare a study on radiation-induced aortitis in a cervical cancer patient | A case study of radiation-induced aortitis | Significant contributions of ChatGPT in preparing the study on radiation-induced aortitis by providing relevant and accurate answers to the questions asked to it |
| (Huh, 2023a) | Korea | Medicine | Assessed and compared ChatGPT's knowledge | Administering a parasitology examination to | Performing better compared to medical students in terms of |
| | | | and inferring skills with medical students | ChatGPT and medical students | knowledge and inferring abilities in a parasitology examination |
| (Gilson et al., 2023) | USA, Ireland | Social Sciences | Examined the performance of ChatGPT in answering questions of the United States Medical Licensing Examination | Two sets of multiple-choice questions from the question banks: AMBOSS and NBME | Identifying the strong potential of ChatGPT as a learning tool for medical education due to its profound knowledge and logical thinking equivalent to a third-year medical student |
| (Lim et al., 2023) | Malaysia, Australia, | Social Sciences, Business, Management and Accounting | Discussed the applications, advantages, and limitations of generative AI in education | Review of existing literature and current events related to generative AI and education | The capabilities of generative AI to transform education promoting innovative use of technology and challenges generative AI faces related to originality of work and plagiarism |
| (Hamdoun et al., 2023) | USA | Social Sciences, Engineering | Explored the implications of digital mental health related applications embedded with chatbots | Analyzing digital mental health applications from a techno-critical perspective | Identifying and discussing digital mental health risks and shortcomings |
| (Lund et al., 2023) | USA, China, India | Social Sciences | Analyzed the impact of ChatGPT on academia and scholarly publishing | An investigation of ethical issues | The huge possibilities of ChatGPT to significantly impact academia and scholarly research, in addition to the ethical issues that need to be considered when adopting this technology |
| (Dowling & Lucey, 2023) | Ireland, UAR, Viet Nam, China | Economics, Econometrics and Finance | Evaluated ChatGPT's effectiveness in assisting finance research | Evaluation of cryptocurrency research outputs using ChatGPT | Underlining the superior performance of ChatGPT on the three versions of finance research ideas discussed in the paper |
| (Kirmani, 2022) | France | Chemistry, Materials Science, Energy | Examined the ability of ChatGPT to write poems | An analysis of the abilities of ChatGPT when it was asked to write a poem perovskite | The remarkable level of diligence, precision, and sophistication showed by ChatGPT in writing a poem by itself. |
| (Mijwil et al., 2023) | Iraq, USA | Computer Science | Discussed the importance of cybersecurity and utilized ChatGPT to determine ways to tackle cybercrime | Generating a section of the article by directly employing ChatGPT | The importance of artificial intelligence in cybersecurity through the writing capabilities of ChatGPT |
| (Selivanov et al., 2023) | Russia | Multidisciplinary | Discussed the automatic clinical image captioning using an integration of two language models | A combined approach comprising of Show-Attend-Tell (SAT) and the GPT-3 | Developing an efficient model in captioning chest X-ray images after testing on two datasets: MIMIC-CXR and MS-COCO |

Several studies have been conducted on using ChatGPT in medicine articles in various aspects. In a study by Nachshon et al., ChatGPT helped to write a description of a case related to 100% third-degree burns (Nachshon et al., 2023). It was found that despite ChatGPT's ability to present facts, it was unable to synthesize or comment on the case's emotional or moral elements. Rockwell et al. examined the diagnosis and treatment of a rare fungal infection where the results were written using ChatGPT (Rockwell et al., 2023). Using the description of simulated datasets, Macdonald et al. evaluated ChatGPT in different aspects, including accelerating paper drafting, generating survival analysis codes, and calculating hazard ratios (Macdonald et al., 2023). The results of the study indicated the usefulness of ChatGPT in the mentioned areas. In another study, the performance of ChatGPT in academic tasks, including summarizing data, suggesting structure, and finding academic papers, was discussed. It was found that ChatGPT could be a useful tool under human supervision (Salvagno, Taccone and Gerli, 2023). The study by Mann on using ChatGPT in analyzing large translational medicine datasets presented the inefficiency of this chatbot due to several limitations such as data dependency, the complexity of biological systems, and ethical implications (Mann, 2023). A case of diagnosing radiation-induced aortitis using imaging studies was reported using ChatGPT in the study by Charrois-Durand et al. (2023). In another study by Huh, the author assessed ChatGPT's ability to answer medical questions as compared to that of students (Huh, 2023a). ChatGPT's accuracy of 60.8% compared to students' accuracy of 90.8% indicated that ChatGPT was not able to perform as well as the medical students.

The application of ChatGPT in social sciences has also attracted significant attention among researchers. The efficiency of ChatGPT compared to other chatbots in the US Medical Licensing Exam using AMBOSS and NBME questions was examined by Gilson et al. (2023), and it was found that ChatGPT outperformed InstructGPT and GPT-3 by 60% and provided logical justifications in 100% of cases. In another study, Lim et al. (2023) evaluated the potential uses and disadvantages of generative artificial intelligence (AI) in education. In addition, in terms of managing mental health, the study conducted by Hamdoun et al. (2023) showed that text-based or voice-enabled conversational agents could be efficient by incorporating ethical considerations. A recent study carried out by Lund et al. (2023) demonstrated the potential for ChatGPT to improve productivity in academia and scholarly publishing.

The use of ChatGPT in other fields has been the subject of several studies, including in finance research, writing abstracts, and cybersecurity. In finance research, Dowling and Lucey (2023) investigated ChatGPT's utility in generating plausible research studies for journals where the output quality was determined by researchers' domain expertise and access to personal data. In another study Kirmani (2022) indicated that ChatGPT’s responses in writing abstracts are promising in terms of presentation, conciseness, and detail sophistication. Furthermore, ChatGPT has contributed to articles on cybersecurity (Mijwil et al., 2023). Additionally, researchers have developed a novel technique for automatic caption generation from radiological scanning data and patient information, which generates interpretable results for clinical decision-making (Selivanov et al., 2023). These studies highlight the potential benefits and challenges associated with the use of generative AI in various fields. In the following, 30 scientific documents on ChatGPT in different fields have been reviewed.

### 3.2.2.2. Other Sources

Recent studies have shown that ChatGPT-generated abstracts can be difficult to distinguish from genuine abstracts, raising concerns about the originality and credibility of scientific papers. There have been several studies conducted on the use of ChatGPT in writing abstracts and collaborating on paper publication. A significant concern is ChatGPT's impact on the writing of scientific papers (González-Padilla, 2023; Gordijn & Have, 2023; Huh, 2023b; Lee, 2023; Patel & Lam, 2023; Selivanov et al., 2023; Yeo-Teh & Tang, 2023). Based on recent studies on assessing the originality of generated abstracts using ChatGPT, it was found that only 63% of fake abstracts were caught by academic reviewers (Thorp, 2023), and in another study (Else, 2023), only 68% of the generated abstracts and 86% of the genuine abstracts were correctly identified. As a result, ChatGPT is able to sneak into academic publications and cannot be held responsible for the content or integrity of scientific papers (Nature Editorial, 2023). Many publishers, authors, and preprint servers agree that this chatbot does not meet the criteria for being a study author. Another reason was that the formal role of an author of a scholarly manuscript must be distinguished from the more general notion of an author as the writer of a document (Stokel-Walker, 2023). Incorporating this chatbot into the academy is problematic. Furthermore, several studies (Flanagin et al., 2023; Hosseini, Rasmussen and Resnik, 2023; Stokel-Walker, 2022; Stokel-Walker & Van Noorden, 2023) have discussed to ChatGPT's erroneous and misleading information, specifically Van Dis et al. (2023) found that ChatGPT generates inaccurate and misleading responses when presented with a series of specific publishing-related technical questions and assignments requiring an in-depth understanding of the literature. In addition, further investigation has revealed that the model was trained on data that covered up until 2021, so it didn't cover more recent events such as COP27 (Chatterjee & Dethlefs, 2023). Despite the problems raised, it is probable that in future, ChatGPT will be developed and will be able to quickly scan the Internet and trawl thousands of openly available scientific articles on different topics to generate a written response with appropriate referencing within milliseconds (O'Connor & ChatGPT, 2023).

On ChatGPT, there are numerous types of non-article scientific documents related to healthcare. ChatGPT can be used to provide medical information and assistance. For instance, ChatGPt can answer medical questions or provide differential diagnoses for common symptoms (Arif, Munaf and Ul-Haque, 2023), gather pertinent clinical information, generate CPT codes from operative reports with a moderate degree of accuracy, and extract ICD-10 codes from clinical notes (DiGiorgio & Ehrenfeld, 2023). A study published in this year investigated applications, such as translating medical language into easier-to-understand text or translating instructions directly into the patient's native language (Scerri & Morin, 2023). ChatGPT can also assist in the development of clinical decision and support systems through the analysis of patient records (Dahmen et al., 2023; De Angelis et al., 2023). Contribution of ChatGPT in diabetes technology was studied by Huang et al. (Huang et al., 2023). In another study, the role of ChatGPT in surgical practice, including anesthesia, and preoperative surgical care, has been studied (Bhattacharya et al., 2023). A further application of ChatGPT in the healthcare system is for generating and improving discharge summaries, which are an obvious choice since their format is largely standardized. Nevertheless, it is important to take into account the fact that the output of this program will need to be reviewed manually by a doctor before it can be completed (Patel & Lam, 2023).

Apart from its applications in writing and healthcare, ChatGPT has gained interest in other areas as well. A study of the potential use of ChatGPT in global warming, which includes the generation of climate scenarios for policy decision-making, and the potential disadvantages, such as lack of contextual awareness and lack of accountability, has been conducted (Biswas, 2023). In another study, ability of ChatGPT to simulate interactions between employees and organizations and help researchers to understand complex organizational dynamics has been evaluated (Dasborough, 2023). Also, the functionality of ChatGPT in tourism and psychiatry has been discussed in several studies (Thornton, D'Souza and Tandon, 2023; Nautiyal et al., 2023). Overall, despite being a relatively new technology, ChatGPT has already found its way into various fields and domains, the most important of which was mentioned and discussed in this paper.

## 4. Discussion

In recent years, chatbots have become increasingly popular due to their versatility in applications such as healthcare, customer service, and education. Among these chatbots, ChatGPT, developed by OpenAI, stands out due to its advanced natural language processing capabilities based on the GPT-3.5 architecture. ChatGPT allows researchers and the general public to ask questions, provide explanations, and engage in conversations about various topics relevant to different fields. Chatbot research has gained interest among researchers worldwide, with participants from around 60% of the world's countries and territories (Figure 3). Despite being a relatively new topic, research on ChatGPT has been conducted in 46 countries (Figure 8). The United States, with its significant research capacity, has the highest number of scientific research institutions and ranks highest in both chatbot and ChatGPT-related publications. Following the United States, Germany and the UK lead in chatbot and ChatGPT research (Figure 3). They are among the top group of countries producing chatbot publications in Scopus and are second and third in the number of ChatGPT publications, respectively.

This study shows that authors in the field of chatbots can be grouped into 15 clusters in both Scopus and Web of Science. Also, authors' impact was compared from different perspectives (Table 4). While Gao J has the most citations in Scopus in chatbot research (1344), FOLSTAD A has the highest number of citations in Web of Science. Additionally, this author has the highest h-index, g-index, m-index, and number of publications in both databases. This author collaborated closely with Araujo, T in Scopus (Figure 4a) and Hobert, S in WoS (Figure 4b).

The analysis of keywords revealed that ChatGPT is a recent trend in the chatbot field, as demonstrated in Figure 7. Furthermore, mental health has become an increasingly popular research area in chatbot development, aided by NLP techniques (Figure 5). It is worth noting that mental health was the leading keyword in its corresponding cluster in both the Scopus and Web of Science databases, as indicated in Tables 5 and 6. In addition, healthcare emerged as a significant area of interest in ChatGPT-related research (Figure 11). Researchers have explored ChatGPT's potential to offer personalized healthcare recommendations, facilitate remote patient monitoring, and assist healthcare professionals in decision-making.

The study has several limitations. Firstly, the research only focused on publications indexed in Scopus and Web of Science, potentially excluding relevant literature from other databases. Secondly, some fields in the records were left empty, which could have impacted the results. These limitations highlight the need for caution when extrapolating the findings and emphasize the importance of conducting additional research to confirm the results.

## 5. Conclusion

This paper employed bibliometric analysis and conducted a thorough literature review on chatbots, with a specific focus on ChatGPT research, spanning from 1998 to 2023, using the Scopus and WoS databases. The research on chatbots is steadily increasing every year, with an annual growth rate of roughly 19% and 27% for WoS and Scopus, respectively. These papers mostly utilize deep learning, machine learning, and NLP techniques. In the past, chatbot research primarily focused on covid-19 and ontology, while the current trend is on mental health and task analysis. ChatGPT is considered a promising research direction in chatbot research, with literature mainly concentrating on its capabilities and limitations in domains such as research ethics, medicine, and social science. The study found that although ChatGPT has some limitations and requires human assistance at times, it has the potential for various applications, including answering medical questions, developing clinical decision and support systems, improving discharge summaries, helping with writing, translating, simulating interactions between employees and organizations, and supporting policy decision-making. As a result, researchers are actively exploring these research directions within the ChatGPT framework.

## Funding

This research did not receive any specific grant from funding agencies in the public, commercial, or not-for-profit sectors.